\documentclass[journal]{IEEEtran}

\normalsize
\usepackage{cite}

\ifCLASSINFOpdf
  \usepackage[pdftex]{graphicx}
  
\else

\fi
  
\begin{document}

\title{Performance Analysis of DF Cooperative Relaying over Bursty Impulsive Noise Channel}

\author{Md Sahabul Alam,~\IEEEmembership{Student Member,~IEEE,}
        ~Fabrice Labeau,~\IEEEmembership{Senior Member,~IEEE},
        and~Georges Kaddoum,~\IEEEmembership{Member,~IEEE}
\thanks{M. S. Alam and G. Kaddoum are with the Electrical Engineering
Department, Ecole de Technologie Superieure, University of Quebec,
Montreal, QC H3C 3P8, Canada (e-mail: md-sahabul.alam.1@ens.etsmtl.ca;
georges.kaddoum@etsmtl.ca).}
\thanks{F. Labeau is with the Electrical and Computer Engineering Department,
McGill University, Montreal, QC H3A 0G4, Canada (e-mail:
fabrice.labeau@mcgill.ca).}}

\markboth{This is the author version of the paper that has been accepted for publication in IEEE Transactions on Communications}%
{Submitted paper}

\maketitle

\begin{abstract}
In this article, we consider the performance analysis of a decode-and-forward (DF) cooperative relaying (CR) scheme over channels impaired by bursty impulsive noise. Although, Middleton class-A model and Bernoulli-Gaussian model give good results to generate a sample distribution of impulsive noise, they fail in replicating the bursty behavior of impulsive noise, as encountered for instance within power substations. To deal with that, we adopt a two-state Markov-Gaussian process for the noise distribution. For this channel, we evaluate the bit error rate (BER) performance of direct transmission (DT) and a DF relaying scheme using M-ary phase shift keying (M-PSK) modulation in the presence of Rayleigh fading with a maximum a posteriori (MAP) receiver. From the obtained results, it is seen that the DF CR scheme in bursty impulsive noise channel still achieves the space diversity and performs significantly better than DT under the same power consumption. Moreover, the proposed MAP receiver attains the lower bound derived for DF CR scheme, and leads to large performance gains compared to the conventional receiving criteria which were optimized for additive white Gaussian noise (AWGN) channel and memoryless impulsive noise channel.
\end{abstract}


\IEEEpeerreviewmaketitle

\section{Introduction}
\IEEEPARstart{T}{he} noise characteristics in many environments, such as around power transmission lines, power substations, and in some mobile radio scenarios are highly non-Gaussian and are inherently impulsive in nature \cite{middleton1977statistical}. For example, in power substations the noise emitted from power equipments, such as transformers, busbars, circuit-breakers, and switch-gears are impulsive \cite{hikita1998electromagnetic,portuguds2003characterisation,sacuto2012evolution}. For smart grid technology \cite{hossain2012smart}, in order to assist the electricity transportation via control, interaction with, and monitoring of power equipment from outside the station, a communication network must be operational within the substation. This could be accomplished by deploying a wireless sensor network (WSN) \cite{gungor2010opportunities,tuna2013wireless} where the deployed sensor nodes collect information from the pieces of equipment, and send their sensed information to the remote smart grid monitoring center for further process. In such applications, the generated impulsive noise from the substation equipment that affects the wireless links between the sensor nodes displays a bursty behaviour as observed in experimental measurements \cite{sacuto2012evolution}.  The models commonly used in the literature to represent impulsive noise are either Middleton class-A \cite{middleton1977statistical} or Bernoulli-Gaussian \cite{ghosh1996analysis}. Although these models give good results to generate a sample distribution of impulsive noise, they cannot describe the bursty nature of the impulses, i.e., the correlation among the noise samples in the time domain. To handle this, Markov chain models have been investigated in the literature \cite{mushkin1989capacity,fertonani2009reliable,ndomarkov,mitra2010convolutionally}, representing the impulsive noise characteristics by including a significant amount of memory.

One of the designing challenges for WSN-based smart grid monitoring applications is how reliably the sensor nodes send their sensed data to the substation monitoring center \cite{gungor2010opportunities,tuna2013wireless}. Cooperative WSNs where the sensor nodes cooperate among each other can be one of the promising candidates for transmission in impulsive channels due to its reliability over fading and interference channels \cite{nosratinia2004cooperative,laneman2004cooperative}. It is based on the broadcast nature of the wireless medium and achieves the potentials of spatial diversity in wireless networks without necessitating the placement of multiple antennas at each node. It is very attractive for WSN since the sensor nodes cannot afford multiple antennas due to their size and cost constraints. The two most popular relaying strategies are DF relaying and amplify-and-forward (AF) relaying. Although there exists a large number of publications on these relaying schemes in various aspects, many of them are restricted to the AWGN assumption. In practice, AWGN is a common assumption to bundle together a lot of sources of noise, beyond thermal. The performance of CR in impulsive channel has only been considered in the literature recently. The pairwise error probability (PEP) of AF CR scheme over flat fading channel in the presence of impulsive noise modeled by Middleton class-A has been investigated in \cite{al2009cooperative,al2009cooperative1}. Upper bounds on PEP expressions are derived for both space time block coded scheme and repetition-based coded scheme. Simulation results demonstrated that the performance of cooperative systems highly depends on the impulsive nature of the noise and different diversity orders are achieved in different signal-to-noise ratio (SNR) regions. Similar performance analysis is carried out in \cite{savoia2011performance} for DF CR schemes. It is shown that similar to the Gaussian noise case, the system achieves full diversity order asymptotically with SNR in impulsive noise scenario. The authors in \cite{siamack2012effect} studied the impact of impulsive noise modeled by a Bernoulli-Gaussian process on the performance of cooperative relaying system in a smart grid scenario. It is shown that as the impulse occurs, probability increases, the performance of the system is getting worse. In \cite{nasri2010performance}, closed-form asymptotic symbol-error rate (SER) and BER expressions were derived for an AF CR scheme with multiple relays which is valid for arbitrary non-Gaussian noise and interference with finite moments. The simulation results reveal that, at high SNR, full diversity order is obtained and is independent of the type of noise. While the above papers quantify the diversity advantages in the presence of impulsive noise, the authors in \cite{van2010performance,van2011effect} studied the performance of DT and DF CR schemes over flat Rayleigh fading and Bernoulli-Gaussian impulsive noise assuming different receiving structures at the destination. The obtained results showed that DF CR performs significantly better than DT under the same bandwidth efficiency and power consumption. It is also shown that while the optimal Bayes receiver \cite{tepedelenlioglu2005diversity} and the maximum ratio combining (MRC) have the same diversity order as expected, the optimal Bayes receiver obtains an additional $3$dB SNR gain over the MRC combiner by considering impulsive noise in the detection process. The above results motivate us to consider the performance analysis of CR over correlated impulsive noise channel. While the complexity of DF relaying is higher than AF relaying duo to its digital processing, we consider DF relaying in our analysis since it reduces the effects of additive noise at the relay \cite{fareed2009relay}.

However, all of the above performance analyses for CR schemes have been carried out over independent and identically distributed (i.i.d.) impulsive channels\footnote{Throughout the article, the terms `impulsive noise' and `impulsive channel' are used interchangeably.}, which cannot include any information on noise time-correlation. To address this issue, we consider a two-state Markov-Gaussian  process \cite{fertonani2009reliable} for noise modeling. A two-state Markov-Gaussian process is a simple and effective way to model a bursty impulsive noise channel \cite{fertonani2009reliable,mitra2010convolutionally}. In this context, the authors in \cite{mushkin1989capacity} calculate the capacity of a Gilbert-Elliott channel which is a varying binary symmetric channel with memory. It is shown that the capacity of the channel increases monotonically with increasing the utilization of memory information at the receiver side and converges to a maximum value which is the capacity of the same channel when perfect state information is available at the receiver. It is also shown that, even if the memory of the channel is ignored through proper interleaving, the capacity of the interleaved channel is lower than the capacity of the original channel. The authors in \cite{fertonani2009reliable} compute the achievable information rate of a two-state Markov-Gaussian channel through an information-theoretic analysis. We would like to point out that, while the state process of a two-state Markov-Gaussian noise model is the same as in a Gilbert-Elliott model, the same analytical arguments do not lead to a closed-form expression for this model since the channel output alphabet is non-binary \cite{fertonani2009reliable,mitra2010convolutionally}. Hence, \cite{fertonani2009reliable} evaluates the information rate of this channel by means of the simulation-based method described in \cite{arnold2006simulation}. It is shown that the ultimate performance limit of such channels improves as the channel memory becomes more significant. Aims at approaching the ultimate performance limit as close as possible, \cite{fertonani2009reliable} provides a transceiver architecture for DT-based on powerful codes and iterative detection. It is shown that the proposed MAP-based iterative receiver with LDPC channel coding is able to exploit the memory of the noise process at the receiver and perform fairly close to the ultimate limit. To the best of our knowledge, no research results have been published on CR schemes impaired by such bursty impulsive channels. Here, we provide a mathematical framework for the performance analysis of DF CR schemes over bursty impulsive noise channel. Our work is an extension of \cite{fertonani2009reliable} to the CR scenario. While we do not attempt to modify the MAP detector proposed in \cite{fertonani2009reliable} to exploit the channel memory, our analysis also includes uncoded scenario and derive analytical error rate expressions for the proposed system, thus providing a framework to validate the simulation results. We expect to gain more compared to the optimal memoryless receiver \cite{van2010performance,tepedelenlioglu2005diversity} proposed for CR scheme over impulsive noise channel by considering noise memory in the detection process. Two different relaying strategies are considered depending on the processing performed by the relay: simple DF relaying (SR) and selective DF relaying (SDFR). In simple DF relaying, the relay transmits all the symbols it receives, whereas, in selective DF relaying, it is assumed that the relay forwards its decoded signal only if the received SNR at the relay is greater than a certain threshold, otherwise the relay remains silent and the destination decodes based on the direct transmission from the source only.

The contributions of this work are as follows. First, we derive a SER formula for DT using M-PSK modulation in the presence of Rayleigh frequency flat fading and two-state Markov-Gaussian impulsive noise. To validate the derived SER formula for DT, we considered the optimal MAP detection criterion that has been used in \cite{fertonani2009reliable} for symbol detection in a two-state Markov-Gaussian noise, and adapt the Bahl-Cocke-Jelinek-Raviv (BCJR) algorithm \cite{bahl1974optimal} to be implemented in the detector for this case. Then, we extend the derived SER formula for DT to the case of DF CR schemes and provide a lower bound under the hypothetical assumption that the receivers have the knowledge of the state of the noise process. Finally, we propose an optimal MAP receiver for the considered DF CR schemes that utilize the MAP detection criterion for each link.

It is shown that the proposed optimal MAP receiver achieves the lower bound derived for DF CR scheme and performs significantly better than the conventional schemes developed for AWGN channel and memoryless impulsive noise channel. Indeed, the BER obtained with the memoryless receiver can be divided by almost $10^{3}$ to get the BER with optimal MAP receiver under coded transmission. Also, DF CR schemes perform significantly better than DT under the same power consumption. In addition, for simple relaying, using the BER of the relay at the destination, the proposed optimal MAP receiver performs significantly better than the case where the MAP receiver does not have any knowledge about the error at the relay and achieves similar performance as that obtained through selective DF relaying.

The rest of the paper is organized as follows. In Section \ref{system_model}, the system model is introduced and Section \ref{two_state_MGM} provides an overview of two-state Markov-Gaussian process. In Section \ref{performance}, we provide the mathematical framework for the proposed scenario. Section \ref{N_results} provides the performances in terms of BER and finally, some conclusions are drawn in Section \ref{conclusion}.

\section{System model}\label{system_model}
\begin{figure}
  \centering
  \includegraphics[scale=0.7]{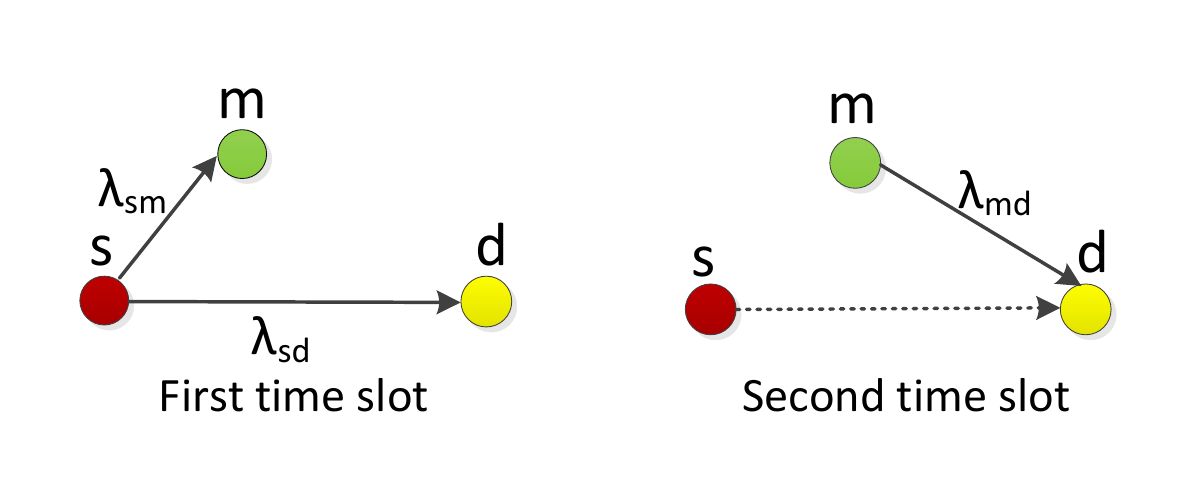}\\
  \caption{Cooperative communication with half-duplex relaying.}\label{relaychannel}
\end{figure}

Here, we consider a DF cooperative relaying scheme with single relay ($m$), as shown in Fig~\ref{relaychannel}, where the data transmission between the source-destination ($sd$) pair is assisted by $m$. We assume that all nodes are equipped with a single antenna and share the same bandwidth for data transmission. We also assume that each node operates in half-duplex mode and hence cannot transmit and receive simultaneously. Both $s$ and $m$ terminals use time division multiplexing for channel access. The cooperative communication takes place in two time slots, with normalized time intervals $t_0$ and $t_1=1-t_0$. In the first time slot, $s$ transmits the data to $d$, and due to the broadcast nature of the wireless channel, $m$ also receives it. The relay then demodulates and decodes the received signal to recover the source information and based on the relaying strategy, it either retransmits in the second time slot of duration $t_1$, or declares that it will remain silent. During this period, $s$ remains in the silent mode as indicated by the dotted line in Fig~\ref{relaychannel}. For simple DF relaying, the relay always retransmits the decoded data to the destination in the second time slot. The destination then receives the noisy observation sequences from $s$ in the first time slot as well as from $m$ in the second time slot. The overall operation is shown in Fig~\ref{relaychannel}. Hence, the decoded data with possible errors are forwarded from the relay to the destination. It is different from most papers on single-relay DF CR schemes where it is typically assumed that if the relay decodes the source message perfectly it will forward its decoded information to the destination, otherwise it will stay in silent mode, i.e., what we call \textit{selective DF relaying} \cite{laneman2004cooperative}, which is decided by comparing the received SNR at the relay to a given threshold. However, in practical relaying systems an arbitrary chosen threshold does not guarantee error-free detection and hence decoding errors may occur at the relay \cite{sneessens2008turbo,lee2009iterative,liang2010relay} even if the received SNR at the relay is greater than a predetermined threshold value. Therefore, though the destination assumes that perfect decoded data were transmitted from the relay, actually the forwarded data may contain hard decision errors. So, there will be a performance degradation if the relay can not be guaranteed to be error-free. Specially, this problem becomes more crucial when the relay moves away from the source and becomes closer to the destination. Although this problem could be solved by considering cyclic redundancy check (CRC) checking at the relays, it is bandwidth-consuming \cite{wang2007high} and induces extensive overhead since CRC checking usually takes place at the MAC layer. The authors in \cite{jayakody2015softforwarding,jayakody2015soft} and the references therein consider soft information relaying to mitigate the effect of decoding error propagation from the relay to the destination. It is shown that soft relaying performs better than hard information relaying under poor source-relay link conditions. However, these schemes require complex offline computation of soft noise parameters which increases the complexity of decoding at the destination for real-time transmission. To avoid these, our analysis remains more general and considers that decoding error might be propagated by the relay.

Consider that the source $s$ generates a frame of binary information of length $L$ bits $(b_0,b_1,\ldots,b_{L-1})$, mapped into a M-PSK modulated sequence $(x_{s,0},x_{s,1},\ldots,x_{s,K-1})$, and transmitted to both $m$ and $d$ in the first time slot. The signals received at $m$ and $d$ at each time epoch $k$, $k=0,1,\ldots,K-1$ can be respectively expressed as
\begin{equation}
y_{{sm,}k}=\sqrt{\textrm{P}_s}h_{sm,k} x_{s,k}+n_{{sm,}k},\label{y_source_relay}
\end{equation}
\begin{equation}
y_{{sd,}k}=\sqrt{\textrm{P}_s}h_{sd,k} x_{s,k}+n_{{sd,}k},\label{y_source_des}
\end{equation}
where $\textrm{P}_s$ is the average source transmission power for each symbol, $x_{s,k}$ is the transmitted symbol from $s$, $h_{ij,k}$ is the channel coefficient for the $ij$ link, $i\in(s,m)$ and $j\in(m,d)$, and $n_{ij,k}$ is the noise term for the $ij$ link that captures the combined effects of AWGN and the impulsive interferers. We assume independent Rayleigh fading in all links, i.e., for each $ij$ link, $h_{ij}\equiv a_{ij}e^{j\theta_{ij}}$ is modeled as a zero-mean, independent, circularly symmetric complex Gaussian random variable with variance $\Omega_{ij}\equiv\varepsilon\{|h_{ij}|^2\}=1/{\lambda_{ij}^\eta}$, where $\varepsilon\{\cdot\}$ denotes expectation operator, $\lambda_{ij}$ is the relative distance of $i$ from $j$, and $\eta$ is the path loss exponent. Hence, the channel amplitudes, $a_{ij}$ are Rayleigh distributed, whereas the channel phases, $\theta_{ij}$ are uniformly distributed in $[-\pi,\pi)$. It is assumed that the channel coefficients are known by the receiver side, but not by the transmitter side. The noise sample $n_{ij,k}$ is modeled as a two-state Markov-Gaussian process, which is in fact a Markov process in with the marginal distribution in each state are Gaussian. In the following section we will provide an overview of the model and explain the physical significance of each parameter. We assume that the noise samples for each link are mutually independent of the other links.

In the second time slot, at $m$, the received signal $y_{sm}$ is passed through a demodulator to recover the source information. The relay then decodes the source information, potentially making an error. After recovering the source information, the relay modulates it using the same modulation format as in $s$ and forwards it to the destination with average transmission power $\textrm{P}_m$. The signal received at the destination in this case is given by
\begin{equation}
y_{{md,}k}=\sqrt{\textrm{P}_m}h_{md,k} x_{m,k}+n_{{md,}k},\label{y_relay_des}
\end{equation}
where $x_{m,k}$ is the transmitted signal from $m$. For fair comparison between DT and CR schemes, in our discussion we assume that the total source transmission power for direct transmission $\textrm{P}_{T}$ is equal to the sum of source and relay transmission power in cooperative communication and hence the total transmission power is constrained as follows:
\begin{equation}
\textrm{P}_{s}+\textrm{P}_{m}=\textrm{P}_{T}.
\end{equation}
\section{An overview of two-state Markov-Gaussian model}\label{two_state_MGM}
\begin{figure}[!t]
  \centering
  \includegraphics[scale=0.8]{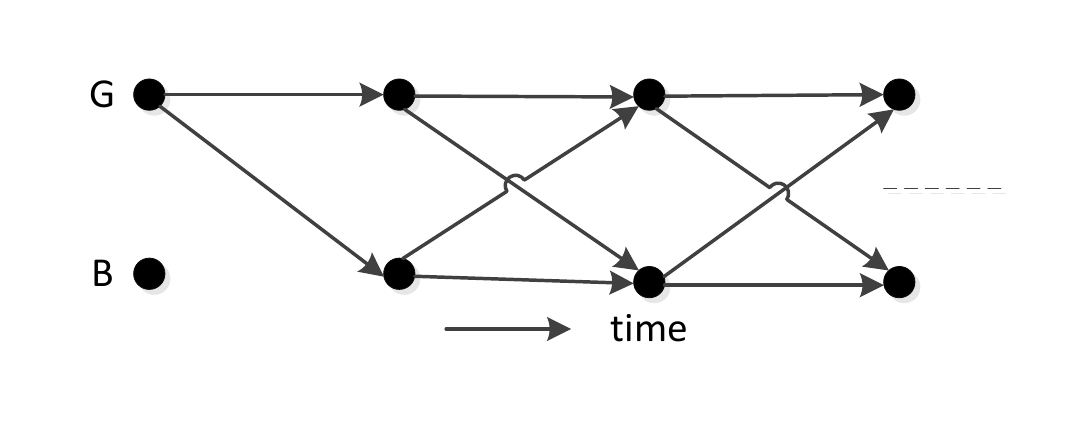}\\
  \caption{Trellis representation of the two-state Markov-Gaussian noise model.}\label{trellisdiagram}
\end{figure}

A two-state Markov-Gaussian model is introduced by Fertonani \cite{fertonani2009reliable} to characterize the correlated impulsive noise. At each time epoch $k$, the statistical properties of the noise sample $n_{ij,k}$ are completely defined by the channel state $s_k \in \{G,B\}$. In our noise modeling, $G$ stands for the good channel that is impaired by the background Gaussian noise only and $B$ for the bad channel which is impaired by impulsive interferers also. For each $ij$ link, we model $n_{ij,k}$ as a zero-mean, circularly symmetric complex Gaussian random variable with variances depending on $s_k$, so that conditioned on $s_k$, the probability density functions (PDFs) of $n_{ij,k}$ can be expressed as
\begin{equation}
 p(n_{ij,k}|s_k\!=\!G)\!\!=\!\!\frac{1}{{\pi \sigma_G^2}} \exp\left(-\frac{|y_{ij,k}\!-\!\sqrt{\textrm{P}_i}h_{ij,k}x_{i,k}|^2}{\sigma_G^2}\right),\label{n_good}
\end{equation}
\begin{equation}
 p(n_{ij,k}|s_k\!\!=\!\!B)\!\!=\!\!\frac{1}{{\pi \sigma_B^2}} \exp\left(-\frac{|y_{ij,k}\!-\!\sqrt{\textrm{P}_i}h_{ij,k}x_{i,k}|^2}{\sigma_B^2}\right),\label{n_bad}
\end{equation}
where $\sigma_G^2$ and $\sigma_B^2$ are the average noise power of the good channel and bad channel respectively. The parameter $R=\sigma_B^2/\sigma_G^2$ quantifies the relative power of the impulsive noise compared to Gaussian noise. The statistical description of the state process $s^K=\{s_0,s_1,\ldots,s_{K-1}\}$ completely characterizes the channel and, for Markov-Gaussian model, $s^K$ is expressed as a stationary first-order Markov-process with
\begin{equation}
p(s^{K})= p(s_0)\prod_{k=0}^{K-1} p(s_{k+1}|s_k),
\end{equation}
for each realization of the process. Therefore, the state process is described by the state transition probabilities $p_{s_ks_{k+1}}=p(s_{k+1}|s_k)$, $s_k,s_{k+1}\in\{G,B\}$. From the state transition probabilities, the stationary probabilities $p_G$ and $p_B$ of being in $G$ and $B$ state are respectively given by \cite{fertonani2009reliable},
\begin{equation}
p_G=p(s_k=G)=\frac{p_{BG}}{p_{GB}+p_{BG}},
\end{equation}
\begin{equation}
p_B=p(s_k=B)=\frac{p_{GB}}{p_{GB}+p_{BG}},
\end{equation}
where $p_{BG}$ denotes the transition probability from state $B$ to state $G$ and similarly $p_{GB}$ is the transition probability from $G$ to $B$. Also, according to the notation in \cite{fertonani2009reliable}, the parameter $\gamma=\frac{1}{p_{GB}+p_{BG}}$ quantifies the noise memory, with $\gamma=1$ meaning that the noise is memoryless and $\gamma>1$ indicating that the noise has persistent memory. Finally, the time evolution of the noise state sequence can be represented by means of a trellis diagram displayed in Fig.~\ref{trellisdiagram}, where all the possible paths given the initial state $G$ are shown. This trellis representation is important for MAP symbol detection and will be discussed in the following section.
\section{Performance analysis}\label{performance}
\subsection{Maximum a Posteriori (MAP) Detection}
For a two-state Markov-Gaussian noise channel, the optimum receiver for DT is designed in \cite{fertonani2009reliable} that exploits the MAP detection criterion for symbol detection. The algorithm derived for MAP symbol detection is based on the factor graphs and the sum-product algorithm assuming no fading. Here, we consider the same detection criterion and summarize the straightforward BCJR algorithm to be implemented into the MAP detector in the presence of Rayleigh fading. The BCJR algorithm is based on the probabilistic arguments and works on a trellis diagram depicted in Fig.~\ref{trellisdiagram} for MAP decoding of two-state Markov-Gaussian noise channel. However instead of a trellis, when it is exported into a Tanner graph, it becomes the sum-product algorithm.

For M-PSK modulation scheme with $M=2$, the MAP decoding rule at the destination is given by
\begin{eqnarray}
\hat{x}_{s,k}{~} = \left\{ \begin{array}{ll}
  1  & \textrm{if $L_k \geq 0$}\\
  -1  & \textrm{if $L_k < 0$}
  \end{array} \right.
\end{eqnarray}
where, $\hat{x}_{s,k}$ is the estimate of the source's transmitted sequence $x_{s,k}$ generated at $d$ and $L_k$ is the log-likelihood ratio (LLR). For direct transmission, $L_k$ at the destination is defined by
\begin{equation}
L_{sd,k}=\ln \left\{ \frac{p(x_{s,k}=1|y_{sd}^K)}{p(x_{s,k}=-1|y_{sd}^K)}\right\}, \label{LLR_relay}
\end{equation}
where $y_{sd}^K=\{y_{sd,0},y_{sd,1},\ldots,y_{sd,K-1}\}$ is the whole sequence to be detected and $K$ is the size of the sequence. For computation, at each $k$, the optimal MAP detector at the destination evaluates the a posteriori probability $p(x_{s,k}|y_{sd}^K)$ for each symbol $x_{s,k}$ belonging to the binary modulation alphabet \{1,-1\}. The a posteriori probability $p(x_{s,k}=b\mid y_{sd}^K)$, $b\in\{1,-1\}$ can be computed from
\begin{equation}
p(x_{s,k}\!=\!b|y_{sd}^K)\!\propto\! p(x_{s,k}\!=\!b,y_{sd}^K)\!=\!\!\!\!\!\!\sum_{s_k,s_{k+1}} \!\!\!\!\!p(x_{s,k}\!=\!b,y_{sd}^K,s_k,s_{k+1})\label{MAP}
\end{equation}
where $s_k,s_{k+1}$ denote the noise states at time $k$ and $k+1$ respectively and the proportionality indicates that the two sides may differ with a positive multiplicative factor that does not have any effect on the detection process \cite{fertonani2009reliable}. Let us define the following quantities
\begin{equation}
\alpha_k(s_k)=p(y_{sd,0},y_{sd,1},\ldots,y_{sd,k-1},s_k),\label{alpha}
\end{equation}
\begin{equation}
\beta_k(s_k)=p(y_{sd,k},y_{sd,k+1},\ldots,y_{sd,K-1}|s_k),\label{beta}
\end{equation}
\begin{equation}
\delta_k(x_{s,k},\!s_k,\!s_{k+1})\!\!=\!\!p(s_{k+1}\!|\!s_k)p(n_{sd,k}\!\!=\!\!y_{sd,k}\!-\!\sqrt{\textrm{P}_s} h_{sd,k} x_{s,k}|s_k)\label{gama}
\end{equation}
where $\alpha_k(s_k)$ and $\beta_k(s_k)$ are referred to as the forward and backward filters and $\delta_k(x_{s,k},s_k,s_{k+1})$ represents the branch metrics of the trellis diagram shown in Fig.~\ref{trellisdiagram}. For a two-state Markov-Gaussian model, the quantity $p(n_{sd,k}=y_{sd,k}-\sqrt{\textrm{P}_s}h_{sd,k}x_{s,k}|s_k)$ is given by (\ref{n_good}) and (\ref{n_bad}) respectively. Assuming independent transmitted symbols $x_{s,k}$, using (\ref{alpha}), (\ref{beta}), and (\ref{gama}), the probability term $p(x_{s,k}=b,y_{sd}^K,s_k,s_{k+1})$ in (\ref{MAP}) can be represented as
\begin{eqnarray}
p(x_{s,k}\!=\!b, y_{sd}^K,s_k,s_{k+1})\!\!\!&=&\!\!\!p(x_{s,k}\!=\!b)\alpha_k(s_k)\beta_{k+1}(s_{k+1})\nonumber\\&&\times\delta_k(x_{s,k}\!=\!b,s_k,s_{k+1}),
\end{eqnarray}
Thus, from (\ref{MAP})
\begin{eqnarray}
p(x_{s,k}\!=\!b,y_{sd}^K)\!=\!p(x_{s,k}\!=\!b)\!\!\!\sum_{s_k,s_{k+1}}\!\!\!\alpha_k(s_k)\beta_{k+1}(s_{k+1})\nonumber\\
\times \delta_k(x_{s,k}\!=\!b,s_k,s_{k+1}),
\end{eqnarray}
Then, the LLR values at the destination are obtained by
\begin{eqnarray}
L_{sd,k}\!=\!\ln \left\{\frac{p(x_{s,k}\!=\!1,y_{sd}^K)}{p(x_{s,k}\!=\!-1,y_{sd}^K)}\right\}\!=\!\ln\big\{\frac{p(x_{s,k}\!=\!1)}{p(x_{s,k}\!=\!-1)}\times \nonumber
\end{eqnarray}
\begin{equation} \frac{\sum_{s_k,s_{k+1}}\alpha_{k}(s_k)\delta_{k}(x_{s,k}\!=\!1,s_k,s_{k+1})\beta_{k+1}(s_{k+1})}{\sum_{s_k,s_{k+1}}\alpha_{k}(s_k)\delta_{k}(x_{s,k}\!=\!-1,s_k,s_{k+1})\beta_{k+1}(s_{k+1})}\}. \label{LLR_finalcase1}
\end{equation}

The forward and backward filters can be computed recursively as
\begin{equation}
\alpha_{k+1}(s_{k+1})=\sum_{s_k,x_{s,k}}\alpha_k(s_k)p(x_{s,k})\delta_k(x_{s,k},s_k,s_{k+1}),
\end{equation}
\begin{equation}
\beta_{k}(s_{k})\!=\!\!\!\sum_{s_{k+1},x_{s,k}}\beta_{k+1}(s_{k+1})p(x_{s,k})\delta_k(x_{s,k},s_k,s_{k+1}),
\end{equation}
where the forward and backward filters are initialized with
\begin{equation}
\alpha_0(s_0=S)=p_S, \: \textrm{and}\: \beta_K(s_K=S)=1. \:  S\in(G,B)
\end{equation}

However, for M-PSK modulation scheme with $M>2$, the first step is to compute the a posteriori probability $p(x_{i,k}|y_{ij,k})$ for each symbol $x_{i,k}$ belonging to the M-PSK modulation alphabet using the MAP symbol detector explained above. The next step is to consider a standard soft demapper \cite{tosato2002simplified} which performs reliable metric computation at the bit level, given the input probability $p(x_{i,k}|y_{ij,k})$. Note that, this block is not needed in case of BPSK since the generated LLR values for the BPSK symbols are the same as the bits. The demapper extracts the quantity $p(b_{l,k}=b|y_{ij,k})$ through the input-output relationship given by
\begin{equation}
p(b_{l,k}=b|y_{ij,k})\propto\sum_{x_{i,k}\in\chi(l,b)}p(x_{i,k}|y_{ij,k}),\label{demapper_MPSK}
\end{equation}
where $b_{l,k}$ is the $l$th bit of symbol $x_{i,k}$ and $\chi(l,b)$ denote the set of $x_{i,k}$ symbols having their $l$th bit equal to $b$. For example, the LLR values for the first and second bit in Q-PSK modulation scheme can be obtained by
\begin{eqnarray}
L_{ij,k}^{(1)}&\!\!\!=\!\!\!&\ln\left\{\frac{b_{1,k}=0|y_{ij,k}}{b_{1,k}=1|y_{ij,k}}\right\}\nonumber\\
&\!\!\!=\!\!\!&\ln\left\{\frac{(x_{i,k}=00|y_{ij,k})+(x_{i,k}=01|y_{ij,k})}{(x_{i,k}=10|y_{ij,k})+(x_{i,k}=11|y_{ij,k})}\right\}.\label{LLR_1}
\end{eqnarray}
\begin{eqnarray}
L_{ij,k}^{(2)}&\!\!\!=\!\!\!&\ln\left\{\frac{b_{2,k}=0|y_{ij,k}}{b_{2,k}=1|y_{ij,k}}\right\}\nonumber\\
&\!\!\!=\!\!\!&\ln\left\{\frac{(x_{i,k}=00|y_{ij,k})+(x_{i,k}=10|y_{ij,k})}{(x_{i,k}=01|y_{ij,k})+(x_{i,k}=11|y_{ij,k})}\right\}.\label{LLR_2}
\end{eqnarray}

It should be mentioned that in order to evaluate the LLR values required for the evaluation of the BER, the receivers need the knowledge of the noise parameters ($p_B,\gamma,R,\sigma_G^2$) and the amplitude of the channel coefficients $h_{ij}$. Similar to \cite{fertonani2009reliable}, it is assumed that these parameters are perfectly known at the receiver side. This assumption is made since we are mainly interested to focus on the BER performance comparison of different receivers, and to evaluate the impact of noise memory. How the receiver side gets these knowledge is beyond the scope of this paper.
\subsection{BER of Direct Transmission}
In order to derive the analytical SER formula for direct transmission, we assume that the destination receiver has the knowledge of the variance of each state. Then, for the considered two-state Markov-Gaussian noise, the conditional probability of symbol error for M-PSK modulated signal when the channel is in good state is given by the integral expression\cite[Eq. (8.23)]{simon2005digital}
\begin{equation}
P_{e,DT}^G=\frac{1}{\pi}\int_{0}^{(M-1)\pi/M} \exp\left(-\frac{\mathrm{P_T}}{\sigma_G^2}\frac{g_{PSK}}{\sin^2\theta}\right),
\end{equation}
where $g_{PSK}=\sin^2(\pi/M)$. Similarly, the conditional SER when the channel is in bad state is given by
\begin{equation}
P_{e,DT}^B=\frac{1}{\pi}\int_{0}^{(M-1)\pi/M} \exp\left(-\frac{\mathrm{P_T}}{\sigma_B^2}\frac{g_{PSK}}{\sin^2\theta}\right),
\end{equation}
Assuming that the receiver has knowledge of the state $s_k$, a lower bound on the average SER for the direct transmission is expressed as
\begin{equation}
P_{e,DT}=p_G^{sd}P_{e,DT}^G+p_B^{sd}P_{e,DT}^B,
\end{equation}
where $p_G^{sd}=p_{BG}^{sd}/p_{GB}^{sd}+p_{BG}^{sd}$ and $p_B^{sd}=p_{GB}^{sd}/p_{GB}^{sd}+p_{BG}^{sd}$ are the steady-state probabilities of having in good state and bad state respectively for the $sd$ link. When fading is present, the conditional SER for a given channel realization $h_{sd}$ is expressed as
\begin{eqnarray}
P_{e,DT}(h_{sd})&\!\!\!\!=\!\!\!\!&\frac{p_G^{sd}}{\pi}\int_{0}^{(M-1)\pi/M} \exp\left(-\gamma_{G}^{sd}\frac{g_{PSK}}{\sin^2\theta}\right)\nonumber\\
&\!\!\!\!+\!\!\!\!&\frac{p_B^{sd}}{\pi}\int_{0}^{(M-1)\pi/M} \exp\left(-\gamma_{B}^{sd}\frac{g_{PSK}}{\sin^2\theta}\right),\label{ber_fade}
\end{eqnarray}
where, $\gamma_{G}^{sd}=\frac{\mathrm{P_T}|h_{sd}|^2}{\sigma_G^2}$ and $\gamma_{B}^{sd}=\frac{\mathrm{P_T}|h_{sd}|^2}{\sigma_B^2}$ are the instantaneous link SNRs for the $sd$ link in good and bad state respectively. Since $h_{sd}\sim CN(0,\Omega_{sd})$, i.e., the link experience Rayleigh fading, $\gamma_{u}^{sd}$ is exponentially distributed with the probability density function
\begin{equation}
f_{\gamma_{u}^{sd}}(\gamma)=\frac{1}{\bar{\gamma}_{u}^{sd}}e^{-\frac{\gamma}{\bar{\gamma}_{u}^{sd}}},
\end{equation}
where, $\bar{\gamma}_{u}^{sd}=\varepsilon\{{\gamma}_{u}^{sd}\}=\frac{\mathrm{P_T}\Omega_{sd}}{\sigma_{u,sd}^2}$ incorporates the average SNR of $sd$ link, $u\in(0,1)\equiv(G,B)$ and $\sigma_{u,sd}^2=R_{sd}^u\sigma_G^2$ is the variance of $n_{sd}$ with $R_{sd}$ is the impulsive to Gaussian noise power ratio for the $sd$ link. By averaging (\ref{ber_fade}) with respect to the random variable $\gamma_{u}^{sd}$ and making use of \cite[Eq. (8.113)]{simon2005digital}, the average SER is given by
\begin{eqnarray}
P_{e,DT}&\!\!\!\!\!\!\!\!=\!\!\!\!\!\!\!\!&\sum_{u=0}^{1}\!\left(\frac{M\!-\!1}{M}\!\right)\!\{1\!-\!\sqrt{\frac{g_{PSK}\bar{\gamma}_{u}^{sd}}{1+g_{PSK}\bar{\gamma}_{u}^{sd}}}\!\left(\frac{M}{(M\!-\!1)\pi}\!\right)\!\!\times\nonumber\\
&&\left[\frac{\pi}{2}\!+\!\tan^{-1}\left(\sqrt{\frac{g_{PSK}\bar{\gamma}_{u}^{sd}}{1+g_{PSK}\bar{\gamma}_{u}^{sd}}}\cot\frac{\pi}{M}\right)\!\!\right]\!\}.
\label{ser_DT}
\end{eqnarray}
For BPSK $(M=2)$, (\ref{ser_DT}) becomes the BER of direct transmission
\begin{equation}
P_{b,DT}=\frac{p_G^{sd}}{2}\left(1-\sqrt{\frac{\bar{\gamma}_{G}^{sd}}{1+\bar{\gamma}_{G}^{sd}}}\right)+\frac{p_B^{sd}}{2}\left(1-\sqrt{\frac{\bar{\gamma}_{B}^{sd}}{1+\bar{\gamma}_{B}^{sd}}}\right).\label{ber_fade_final_DT}
\end{equation}
\subsection{BER of DF Cooperative Relaying}
In case of DF cooperative relaying, the SER at the relay follows the same form as in (\ref{ser_DT}), i.e.,
\begin{eqnarray}
P_{e,m}&\!\!\!\!\!\!\!\!=\!\!\!\!\!\!\!\!&\sum_{u=0}^{1}\!\left(\frac{M\!-\!1}{M}\!\right)\!\{1\!-\!\sqrt{\frac{g_{PSK}\bar{\gamma}_{u}^{sm}}{1+g_{PSK}\bar{\gamma}_{u}^{sm}}}\!\left(\frac{M}{(M\!-\!1)\pi}\!\right)\!\!\times\nonumber\\
&&\left[\frac{\pi}{2}\!+\!\tan^{-1}\left(\sqrt{\frac{g_{PSK}\bar{\gamma}_{u}^{sm}}{1+g_{PSK}\bar{\gamma}_{u}^{sm}}}\cot\frac{\pi}{M}\right)\!\!\right]\!\}.
\label{ser_relay}
\end{eqnarray}
where, $\bar{\gamma}_{u}^{sm}=\frac{\mathrm{P_T}\Omega_{sm}}{R_{sm}^u\sigma_G^2}$, $u\in\{G,B\}$ is the average received SNR of $sm$ link. The end-to-end SER performance of DF cooperative relaying scheme depends on different relaying strategies such as SR in which the relay always transmits in the second phase. The end-to-end SER under this condition is equal to
\begin{equation}
P_{e,coop}^{SR}= P_{e,m}\cdot P_{e,smd}^{er}+ (1-P_{e,m})\cdot P_{e,smd}^{ner},\label{ber_theory_simple}
\end{equation}
where $P_{e,smd}^{er}$ is the probability of error at the destination after combining the two signals coming from the source and the relay when the error is propagated from the relay. Also, $P_{e,smd}^{ner}$ is the probability of error at the destination when there is no error propagation from the relay and hence the source and the relay will transmit the same data. On the other hand, in SDFR it is assumed that the relay forwards its decoded signal only if the source-relay SNR is larger than a certain threshold, otherwise the relay remains silent and the destination decodes based on the direct transmission from the source only. A lower bound of this protocol is obtained if it is assumed that the relay is able to decode the source symbol successfully when the received SNR at the relay is greater than a certain threshold and retransmits on the second phase only if it is successfully decoded. The average SER at the destination under this scheme can be computed as
\begin{equation}
P_{e,coop}^{SDFR}= P_{e,m}\cdot P_{e,DT}+(1-P_{e,m})\cdot P_{e,smd}^{ner},\label{ber_theory_ideal}
\end{equation}
However, in practical relaying systems, an arbitrary chosen threshold does not guarantee error-free detection, and hence decoding errors may occur at the relay even if the received SNR at the relay is greater than a predetermined threshold value. The actual average SER at the destination for this scheme for a given threshold $\gamma_t$ can be expressed as
\begin{eqnarray}
P_{e,coop}^{SDFR}\!=\!p\left\{\gamma_{u}^{sm}\!>\!\gamma_t \right\}[P_{e,m}|\gamma_{u}^{sm}\!>\!\gamma_t \cdot P_{e,smd}^{er}+\nonumber
\end{eqnarray}
\begin{equation}
(1-P_{e,m}|\gamma_{u}^{sm}>\gamma_t)\cdot P_{e,smd}^{ner}]+p\left\{\gamma_{u}^{sm}\leq\gamma_t \right\}\cdot P_{e,DT},\label{ber_SDF}
\end{equation}
Since $\gamma_{u}^{sm}$ is an exponential random variable with mean $\bar{\gamma}_{u}^{sm}$, we have
\begin{equation}
p\{\gamma_{u}^{sm}\leq \gamma_t\}=1-\exp\left(-\gamma_t/\bar{\gamma}_{u}^{sm}\right),
\end{equation}
When $\gamma_{u}^{sm}>\gamma_t$, the SER at the relay decreases, but it remains nonzero regardless of the value of $\gamma_t$ \cite{onat2008threshold}. Following the same procedure in \cite{onat2008threshold}, for BPSK/Q-PSK modulation scheme the BER at the relay given that $\gamma_{u}^{sm}>\gamma_t$ is equal to
\begin{eqnarray}
P_{b,m}|\gamma_{u}^{sm}\!>\!\gamma_t=\frac{1}{2}\sum_{u=0}^{1}(p_B^{sm})^u(p_G^{sm})^{1-u}[erfc(\sqrt{\gamma_t})\nonumber\\
-e^{\gamma_t/\bar{\gamma}_{u}^{sm}}\sqrt{\frac{\bar{\gamma}_{u}^{sm}}{1+\bar{\gamma}_{u}^{sm}}} erfc\left(\sqrt{\gamma_t\left(1+\frac{1}{\bar{\gamma}_{u}^{sm}}\right)}\right)].
\end{eqnarray}

\begin{figure}
  \centering
  \includegraphics[scale=0.6]{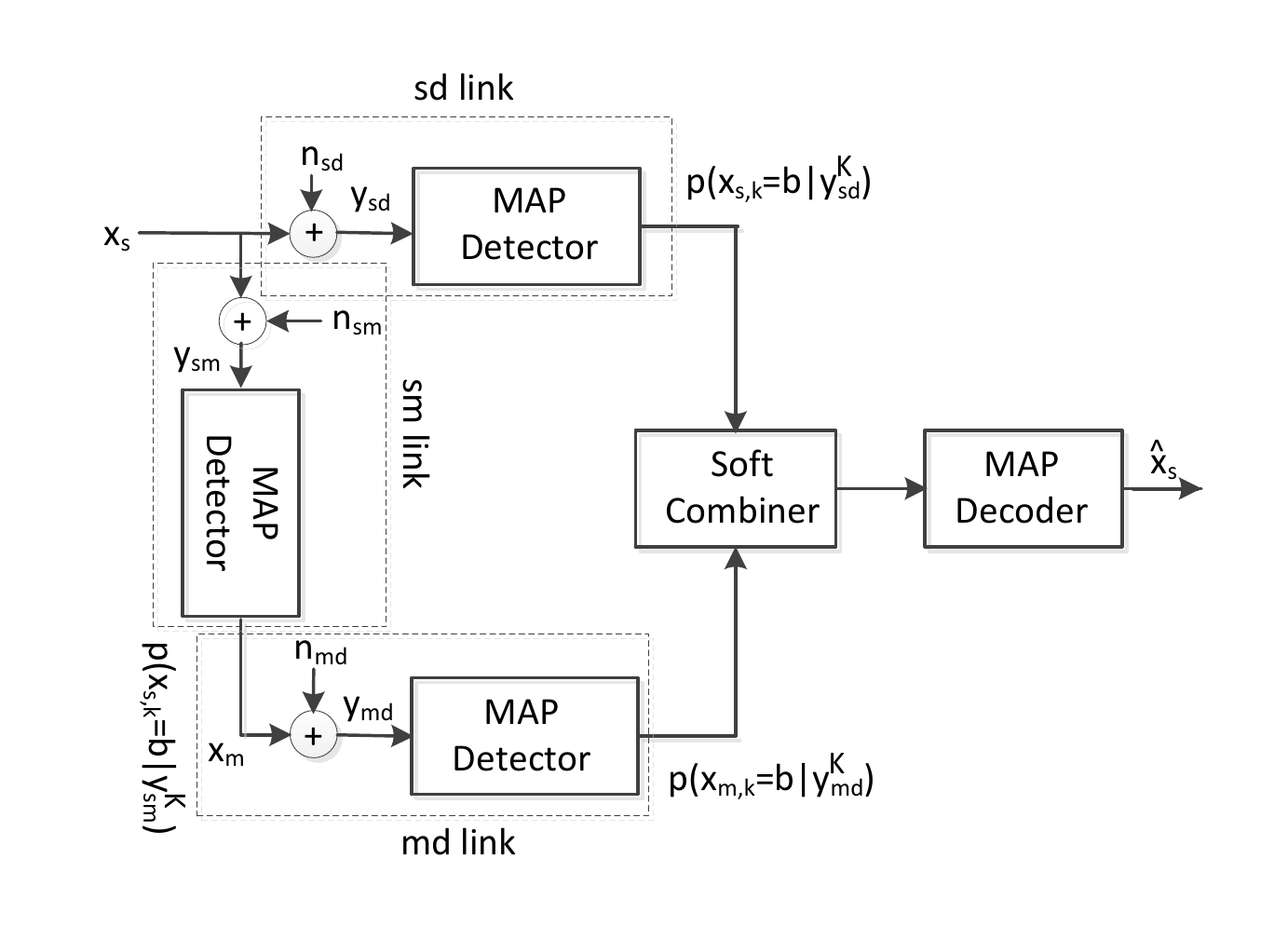}\\
  \caption{MAP receiver for DF cooperative relaying over correlated impulsive noise channel. The system is composed of three MAP detectors, one for each link.}\label{receiver}
\end{figure}

In order to compute $P_{e,smd}^{er}$ and $P_{e,smd}^{ner}$ we have to know which combiner is used for combining the signals coming from the source and the relay. For AWGN channel, i.e., when impulsive noise is absent, the maximum ratio combining is optimal in the sense of minimizing the SER. The MRC combining is
\begin{equation}
y_d=\sqrt{\mathrm{P_s}}h_{sd}^*y_{sd}+\sqrt{\mathrm{P_m}}h_{md}^*y_{md}.\label{MRC_combining}
\end{equation}
When impulsive noise is present, the optimal MAP combining is
\begin{eqnarray}
L_{coop,k}=\ln \left\{\frac{p(x_{s,k}=x_0|y_{sd}^K,y_{md}^K)}{ p(x_{s,k}=x_z|y_{sd}^K, y_{md}^K)}\right\}\nonumber
\end{eqnarray}
\begin{equation}
=\ln \left\{\frac{\sum\limits_{x_{m,k}\in \{x_0,\ldots,x_z,\ldots,x_{M-1}\}}p(x_{s,k}=x_0, x_{m,k}|y_{sd}^K,y_{md}^K)}{\sum\limits_{x_{m,k}\in \{x_0,\ldots,x_z,\ldots,x_{M-1}\}}p(x_{s,k}=x_z, x_{m,k}|y_{sd}^K, y_{md}^K)}\right\},\label{LLR_coop}
\end{equation}
where $L_{coop,k}$ is the symbol LLR value at the destination in case of cooperative communication. Without loss of generality, it is assumed that the source transmits $x_0$. Using the Bayes rule, the following probability term is given by
\begin{equation}
p(x_{s,k},x_{m,k}|y_{sd}^K,y_{md}^K)=\frac {p(y_{sd}^K,y_{md}^K|x_{s,k},x_{m,k})p(x_{s,k},x_{m,k})}{p(y_{sd}^K,y_{md}^K)},
\end{equation}
We assume that for the given transmitted signals, $y_{sd}^K$ and $y_{md}^K$ are independent from each other. Under this consideration, we have
\begin{eqnarray}
p(y_{sd}^K,y_{md}^K|x_{s,k},x_{m,k})= p(y_{sd}^K|x_{s,k})\cdot p(y_{md}^K|x_{m,k}),\label{independence}
\end{eqnarray}
where the equality is due to the conditional independence of $y_{sd}^K$ and $y_{md}^K$ and using the facts that $p(y_{sd}^K|x_{s,k},x_{m,k})\!=\!p(y_{sd}^K|x_{s,k})$ and $p(y_{md}^K|x_{s,k},x_{m,k})\!=\!p(y_{md}^K|x_{m,k})$. Substituting (\ref{independence}) in (\ref{LLR_coop}), results in
\begin{eqnarray}
L_{coop,k}\!\!\!\!&=&\!\!\!\!\ln \left\{\!\frac{p(x_{s,k}\!=\!x_0,y_{sd}^K)}{p(x_{s,k}\!=\!x_z,y_{sd}^K)}\right\}\!+\!\ln \left\{\!\frac{p(x_{m,k}\!=\!x_0,y_{md}^K)}{p(x_{m,k}\!=\!x_z,y_{md}^K)}\right\}\nonumber\\
&&+\ln \left\{\frac{1+\frac{q_m}{1-q_m}(\frac{p(x_{m,k}=x_z,y_{md}^K)}{p(x_{m,k}=x_0,y_{md}^K)})}{1+\frac{q_m}{1-q_m}(\frac{p(x_{m,k}=x_0,y_{md}^K)}{p(x_{m,k}=x_z,y_{md}^K)})}\right\}.\label{LLD_destination_symbol}
\end{eqnarray}
where $q_m=\sum_{z=1}^{M-1}p(x_{m,k}=x_z|x_{s,k}=x_0)$ is the SER at the relay. With M-PSK modulation, there are $M-1$ ways of making an incorrect decision at the relay and their impacts on the detection process at the destination should be different. For BPSK modulation scheme, (\ref{LLD_destination_symbol}) reduces to
\begin{eqnarray}
L_{coop,k}\!\!\!\!&=&\!\!\!\!\ln \left\{\!\frac{p(x_{s,k}\!=\!1,y_{sd}^K)}{p(x_{s,k}\!=\!-1,y_{sd}^K)}\right\}\!+\!\ln \left\{\!\frac{p(x_{m,k}\!=\!1,y_{md}^K)}{p(x_{m,k}\!=\!-1,y_{md}^K)}\right\}\nonumber\\
&&+\ln \left\{\frac{1\!+\!\frac{\theta_m}{1-\theta_m}(\frac{p(x_{m,k}\!=\!-1,y_{md}^K)}{p(x_{m,k}\!=\!1,y_{md}^K)})}{1\!+\!\frac{\theta_m}{1-\theta_m}(\frac{p(x_{m,k}\!=\!1,y_{md}^K)}{p(x_{m,k}\!=\!-1,y_{md}^K)})}\right\}.\label{LLD_destination}
\end{eqnarray}
where $\theta_m$ be the average probability of bit error in detecting the source information at $m$.\footnote{There is of course an underlying assumption here that the average on probability is the same as the total average.} The second term in (\ref{LLD_destination}) can be computed as (\ref{LLR_finalcase1}) with the computation of $\alpha$, $\beta$, and $\delta$ for the $md$ link. The receiver at the destination is then composed of two MAP detectors, one for detecting the source transmission and the other for the relay's transmission. The third term can be estimated by knowing the average error probability at the relay ($\theta_m$) and the output of the MAP detector for the $md$ link. The soft information  are then combined using a soft combiner and input to the MAP decoder to regenerate the information bits. The whole operation is shown in Fig.~\ref{receiver}. We assume that in addition to the decoded bits, the relay also transmits to the destination some side information, for example, the relay may transmit the value of the channel state information \cite{sneessens2008turbo}, so that the decoder at the destination can determine the corresponding error probabilities in the relayed signal.

In order to derive the SER, it is assumed that the destination receiver has the knowledge of the states of $n_{sd}$ and $n_{md}$, and variances of each state. This makes the problem tractable and constitutes a lower bound on the actual SER. Under this consideration, the optimal combiner is based on MRC \cite{van2010performance}. For MRC, the SNR after combining the two signals is the sum of the SNRs of the $sd$ and $md$ links and conditioned on ${\sigma}^2=[\sigma_{u,sd}^2 \quad \sigma_{v,md}^2]$, $P_{e,smd}^{ner}$ is the SER of a two-branch MRC receiver in Rayleigh fading channel which is given in \cite[Eq. (9.14)]{simon2005digital}. For BPSK modulation with independent and non-identically distributed (i.n.d.) Rayleigh channels, this is given as \cite[Eq. (14.5-28)]{proakis2001digital}
\begin{equation}
P_{b,smd}^{ner}(\sigma_{u,sd}^2,\sigma_{v,md}^2)=\frac{1}{2}\left(\frac{\psi(\bar{\gamma}_{u}^{sd})}{1-\bar{\gamma}_{v}^{md}/\bar{\gamma}_{u}^{sd}}+\frac{\psi(\bar{\gamma}_{v}^{md})}{1-\bar{\gamma}_{u}^{sd}/\bar{\gamma}_{v}^{md}}\right),
\end{equation}
where $\psi(\bar{\gamma})=1-\sqrt{\frac{\bar{\gamma}}{1+\bar{\gamma}}}$ and $\sigma_{u,sd}^2=R_{sd}^u\sigma_G^2$ and $\sigma_{v,md}^2=R_{md}^v\sigma_G^2$ are the variances of $n_{sd}$ and $n_{md}$, respectively. By averaging $P_{b-smd}^{ner}(\sigma_{u,sd}^2,\sigma_{v,md}^2)$ with respect to $\sigma_{u,sd}^2$ and $\sigma_{v,md}^2$, we obtain the average BER for the $smd$ link as
\begin{eqnarray}
P_{b,smd}^{ner}=\frac{1}{2}\sum_{u=0}^{1}\sum_{v=0}^{1}(p_B^{sd})^u(p_G^{sd})^{1-u}(p_B^{md})^v(p_G^{md})^{1-v}\times\nonumber\\
(\frac{\psi(\bar{\gamma}_{u}^{sd})}{1-\bar{\gamma}_{v}^{md}/\bar{\gamma}_{u}^{sd}}+\frac{\psi(\bar{\gamma}_{v}^{md})}{1-\bar{\gamma}_{u}^{sd}/\bar{\gamma}_{v}^{md}}).
\end{eqnarray}
To calculate $P_{e,smd}^{er}$, similar to \cite{onat2008threshold}, it is assumed that the dominant cause of detection errors at the destination is due to the incorrectly detected symbol error sent by the relay. For Rayleigh faded channel, in the absence of impulsive noise the error probability under this condition can be approximated by \cite{onat2008threshold}
\begin{equation}
P_{e,smd}^{er}(\sigma_{u,sd}^2,\sigma_{v,md}^2)=\frac{\bar{\gamma}_{v}^{md}C_{z,M}}{\bar{\gamma}_{v}^{md}C_{z,M}+\bar{\gamma}_{u}^{sd}},
\end{equation}
where $C_{z,M}$ depends on the particular value of $M$ and is defined in \cite{onat2008threshold}. In the special case of $M=2$, $C_{z,M}=1$. Then, for impulsive noise channel the average BER for the $smd$ link becomes
\begin{equation}
P_{b,smd}^{er}\!\!=\!\!\sum_{u=0}^{1}\!\sum_{v=0}^{1}\!(p_B^{sd})^u(p_G^{sd})^{1-u}(p_B^{md})^v(p_G^{md})^{1-v} (\frac{\bar{\gamma}_{v}^{md}}{\bar{\gamma}_{v}^{md}+\bar{\gamma}_{u}^{sd}}).
\end{equation}
\begin{figure}[!t]
  \centering
  \includegraphics[scale=0.31]{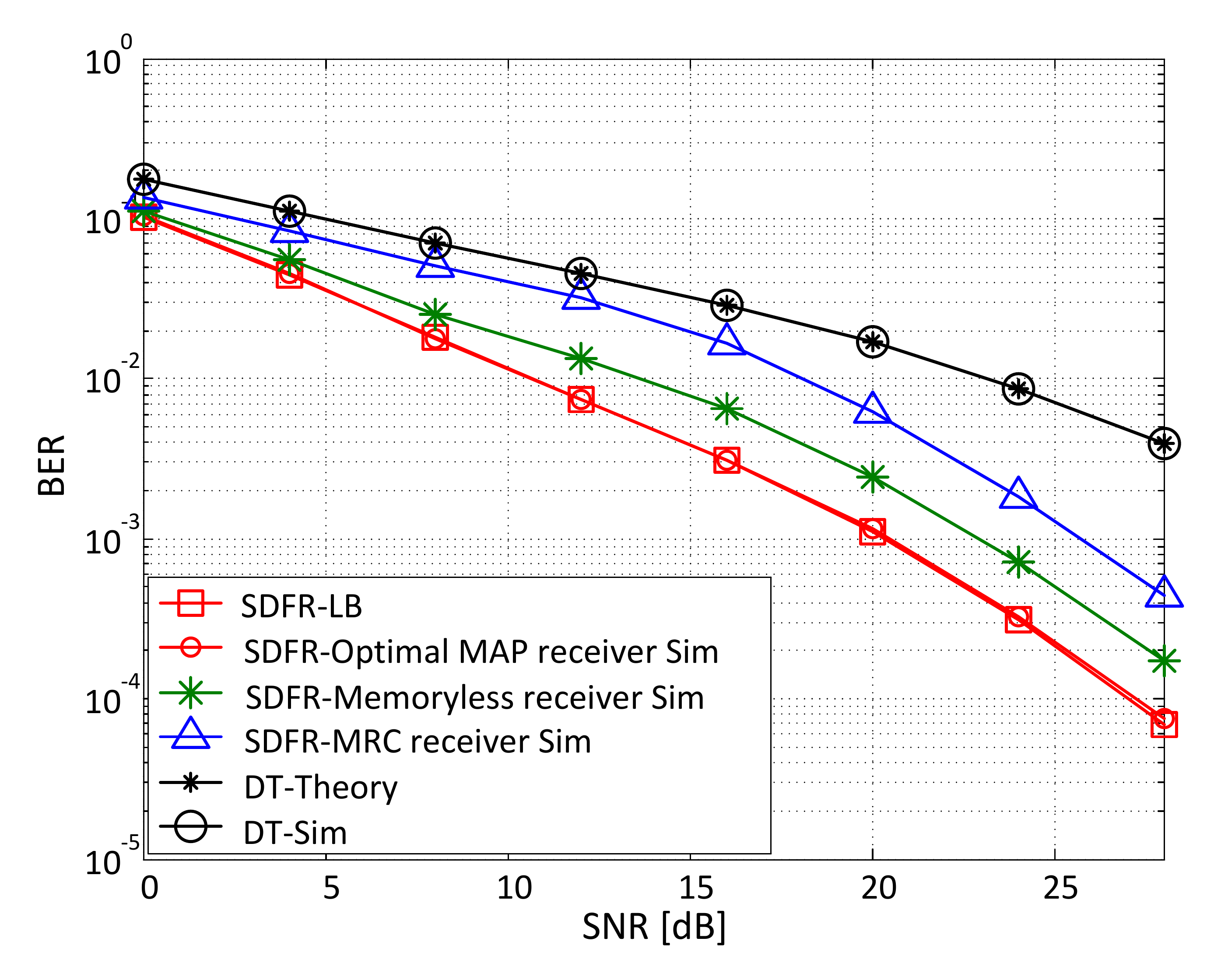}\\
  \caption{Analytical and simulated BER performances of direct transmission (DT) and selection decode-and-forward relaying (SDFR) schemes against SNR. A system employing a BPSK modulation is considered and the performance of various decoding schemes over two-state Markov-Gaussian channels, each characterized by $p_B=0.1$, $\gamma=100$, $R=100$ is shown.}\label{ber_SDFR}
\end{figure}

\section{Numerical results}\label{N_results}
Here, first we present the BER performances of DT and DF CR schemes where a sequence of equally likely information bits of length $64,800$ is mapped onto BPSK modulation sequence and transmitted over two state Markov-Gaussian channels characterized by the identical parameters of bad state occurring rate $p_B=0.1$, channel memory $\gamma=100$, and impulsive to Gaussian noise power ratio $R=100$. In our simulations, it is assumed that the distance between the source and the destination is normalized to unity, i.e., $\lambda_{sd}=1$ and $\lambda_{sr}=0.4$, $\lambda_{rd}=0.6$. Also, slot durations $t_0=t_1=1/2$, both the source and the relay transmit power $\textrm{P}_s=\textrm{P}_m=\textrm{P}_{T}/2$, and the path loss exponent $\eta=2$.
\begin{figure}[!t]
  \centering
  \includegraphics[scale=0.31]{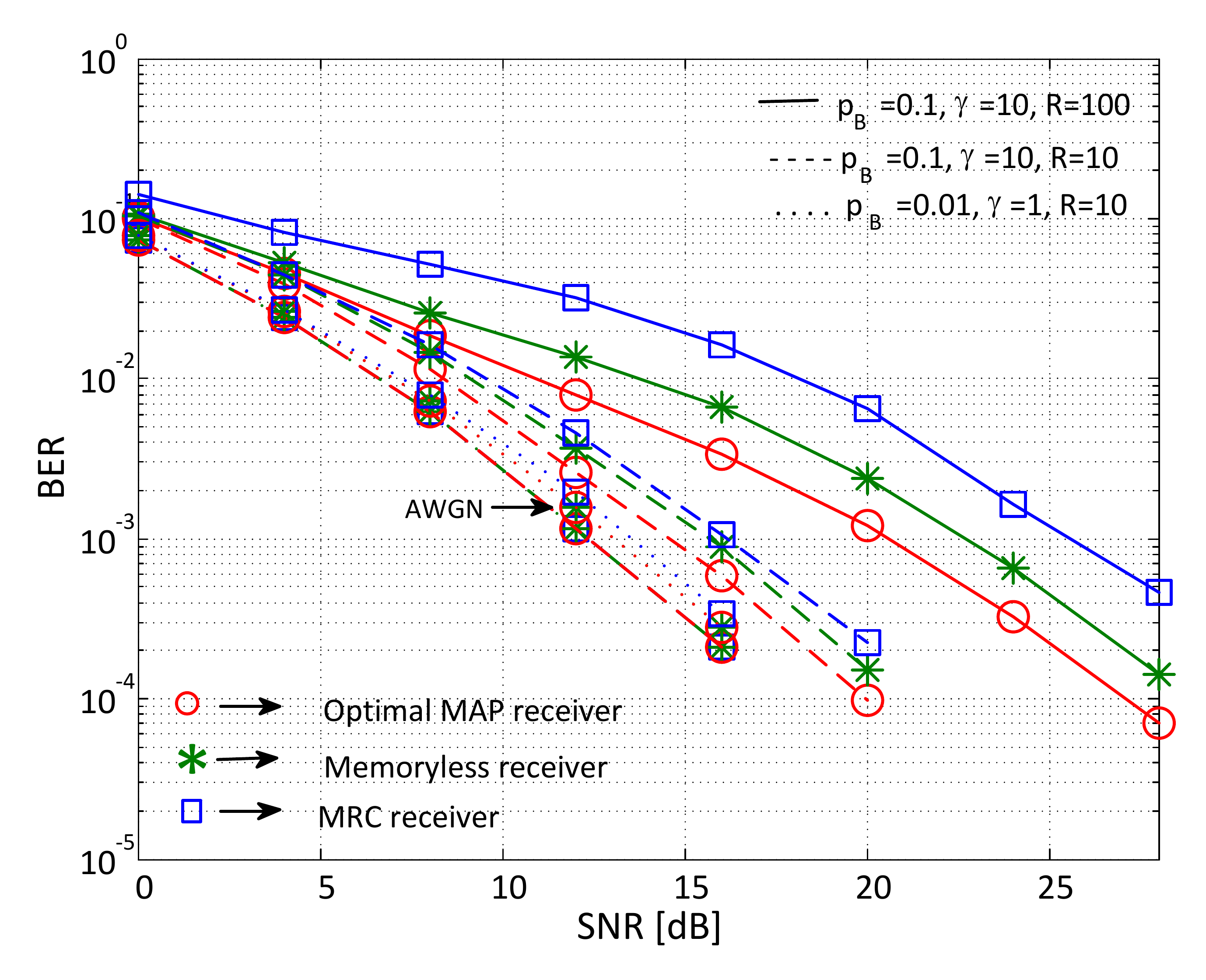}\\
  \caption{BER performances of selection decode-and-forward relaying (SDFR) scheme. A BPSK modulation is adopted and the effect of various noise parameters are considered.}\label{ber_parameter_effect}
\end{figure}

Fig.~\ref{ber_SDFR} shows the analytical and simulated BER performances of both DT and SDFR schemes assuming different receiver structures. The proposed optimal MAP receiver uses the MAP detection criterion, the memoryless receiver \cite{tepedelenlioglu2005diversity} is optimal for i.i.d. Bernoulli-Gaussian noise, and the MRC combiner \cite{proakis2001digital} is optimal for AWGN channel. Similar to \cite{proakis2001digital,van2010performance}, it is assumed that the relay is able to detect whether the source symbol is correctly detected or not, and that it forwards to the destination only if it is correctly decoded.  The exact BER expression in (\ref{ber_fade_final_DT}) and the lower bound of BER expression in (\ref{ber_theory_ideal}) are used to obtain the analytical results for DT and SDFR, respectively. For the simulation results, it is assumed that the noise samples $n_{sm}$, $n_{sd}$, and $n_{md}$ are mutually independent, with each characterized by the noise parameter values listed above. The BER performances are calculated for $2000$ frames of $64,800$ information bits each against $SNR$. The $SNR$ is defined as, $SNR=\varepsilon\{|x_{s,k}|^2\}/\sigma_G^2$, where $\sigma_G^2$ is the background Gaussian noise power. For the considered BPSK modulated signal, $\varepsilon\{|x_{s,k}|^2\}$ is equal to one and the Gaussian noise power $\sigma_G^2$ is adjusted to achieve a given SNR. Also, the SNR is equal to the SNR of the $sd$ link, because the distance between $s$ and $d$ is normalized to unity. To obtain the simulated BER, the LLR values for the direct transmission from source-to-relay and source-to-destination links are obtained using the formula in (\ref{LLR_finalcase1}), and the LLR values for the cooperative $smd$ link are obtained using (\ref{LLD_destination}) with the assumption that $x_{s,k}=x_{m,k}$ and hence $\theta_m=0$. From Fig.~\ref{ber_SDFR}, it is seen that the analytical result perfectly matches with the simulation result for DT and SDFR schemes. Also, SDFR performs significantly better than DT under the same power consumption which confirms the benefit of utilizing CR over bursty impulsive noise channel. Moreover, our proposed MAP receiver achieves the lower bound derived for SDFR. It obtains a minimum SNR gain of around $5$ dB over the MRC combiner in (\ref{MRC_combining}) and around $2$ dB over the optimal memoryless receiver at the expense of higher complexity due to the MAP detection. This confirms the benefits of utilizing the noise memory in the detection process.
\begin{figure}[!t]
  \centering
  \includegraphics[scale=0.31]{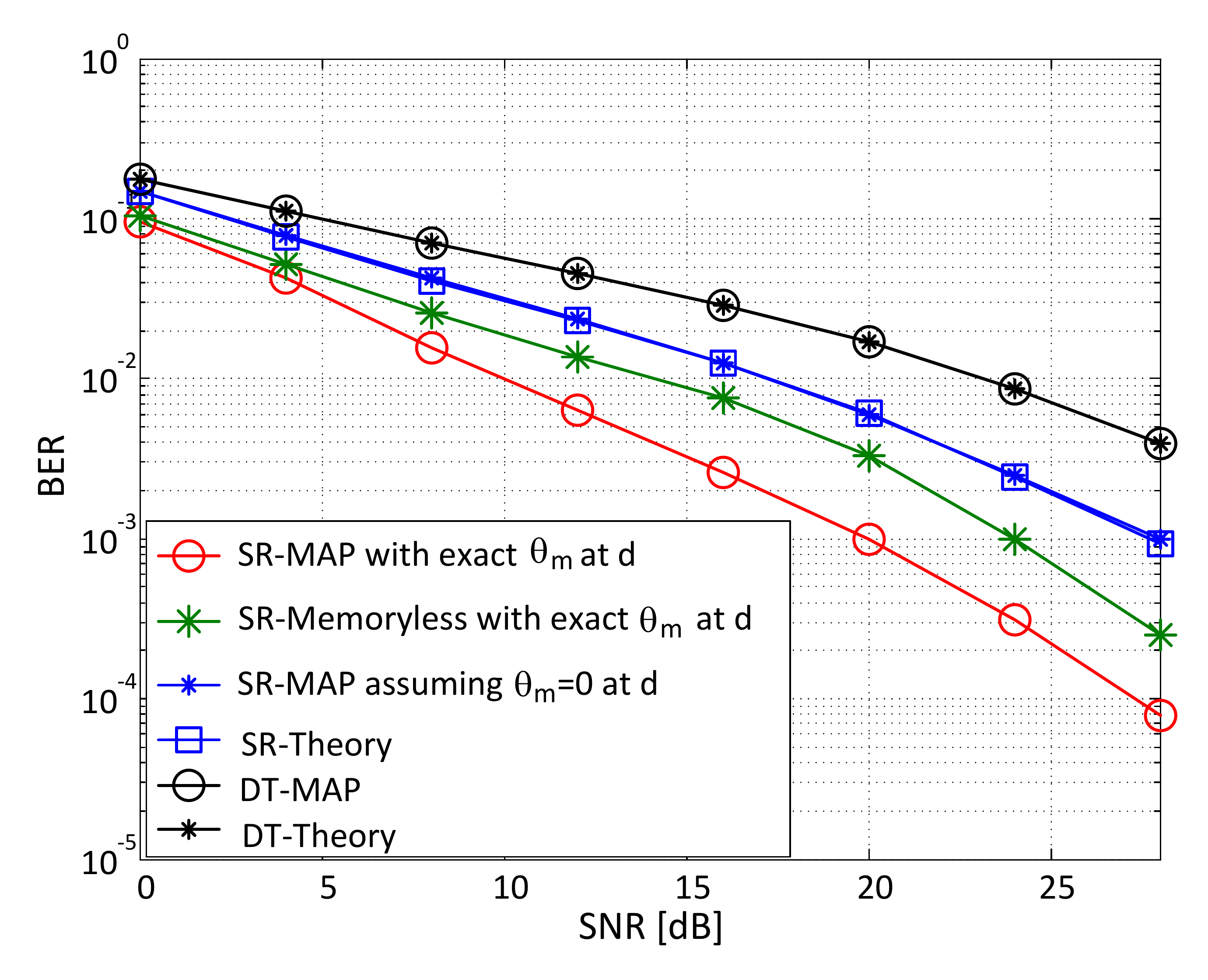}\\
  \caption{Analytical and simulated BER performances of direct transmission (DT) and simple relaying (SR) schemes against SNR with different realizations of $\theta_m$ at the destination. A BPSK modulation is adopted and each channel is characterized by $p_B=0.1$, $\gamma=100$, $R=100$.}\label{ber_SR}
\end{figure}

Although similar conclusions hold for all values of $p_B$, $\gamma$, and $R$, the performance gain provided by the utilization of memory in the detection process depends on those values. This is shown in Fig.~\ref{ber_parameter_effect} for different realizations of $p_B$, $\gamma$, and $R$. From the figure it is seen that for a given value of $p_B$, as the value of $R$ increases, the BER performance degrades. Interestingly, from the figure it is also seen that with increasing $R$, the gain provided by the memory increases. This implies that the larger the value of the impulsive interferers are, the better the performance gain provided by the memory. Also, the optimal MAP receiver shows the same performance as the memoryless receiver when we consider $\gamma=1$ in the noise process, which corresponds to the i.i.d. case of Markov-Gaussian noise commonly known as Bernoulli-Gaussian noise in the literature. This is expected since the memoryless receiver is optimal for i.i.d. Bernoulli-Gaussian noise. These results confirm that the optimal MAP receiver reduces to the memoryless receiver when there is no time correlation among the noise samples. Again, the AWGN receiver achieves the worst performances in these impulsive environments. Finally, we also reported the corresponding curves for an AWGN channel. From the obtained results it is obvious that the three receivers show the same performance over AWGN channel.
\begin{figure}[!t]
  \centering
  \includegraphics[scale=0.31]{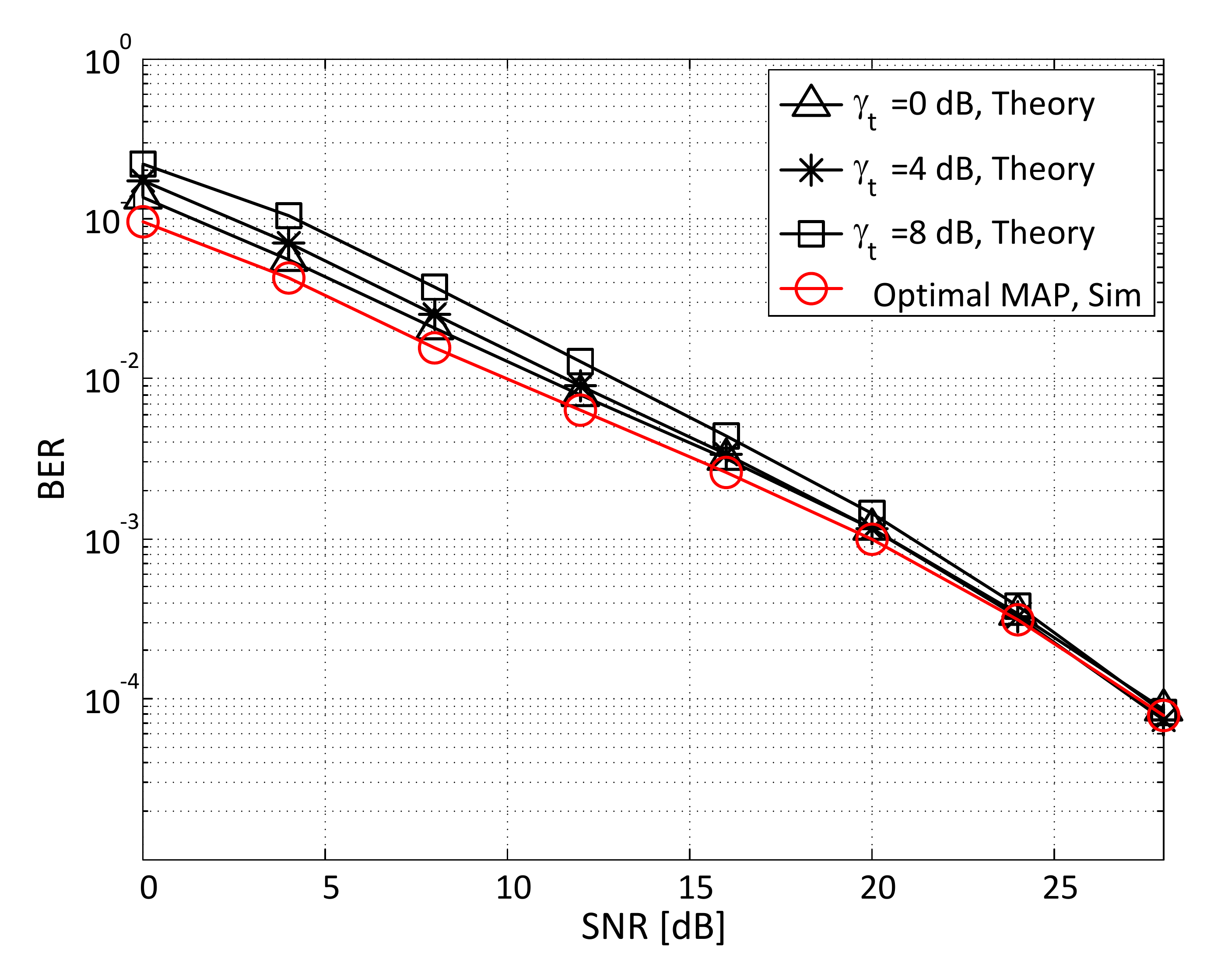}\\
  \caption{BER performances of threshold-based selection decode-and-forward relaying (SDFR) scheme with different values of threshold $\gamma_t$. A BPSK modulation is adopted and each channel is characterized by $p_B=0.1$, $\gamma=10$, $R=10$.}\label{ber_selection_relaying}
\end{figure}

Fig.~\ref{ber_SR} compares the BER performances of DT with SR scheme. The analytical BER for SR scheme is obtained using the formula in (\ref{ber_theory_simple}). For the simulation results, the following cases are considered: (i)- the destination has perfect knowledge about $\theta_m$, which is utilized in the detection process using (\ref{LLD_destination}) and (ii)- when $\theta_m$ is not utilized by the destination, the LLR values are obtained using the first two terms of (\ref{LLD_destination}). From Fig.~\ref{ber_SR} it is seen that the simulation result obtained in case (ii) perfectly matches the analytical result. In addition, our proposed optimal MAP decoder in case (i) showed better performance than in case (ii). It achieves a SNR gain of around $8$ dB by exploiting $\theta_m$ at the destination. Also, SR scheme performs significantly better than DT under both cases. From Fig.~\ref{ber_SR} it is further verified that similar to SDFR, in case of SR, the optimal MAP receiver performs significantly better than the optimal memoryless receiver \cite{huynh2012improved} when both utilizes $\theta_m$ at the destination.
\begin{figure}[!t]
  \centering
  \includegraphics[scale=0.31]{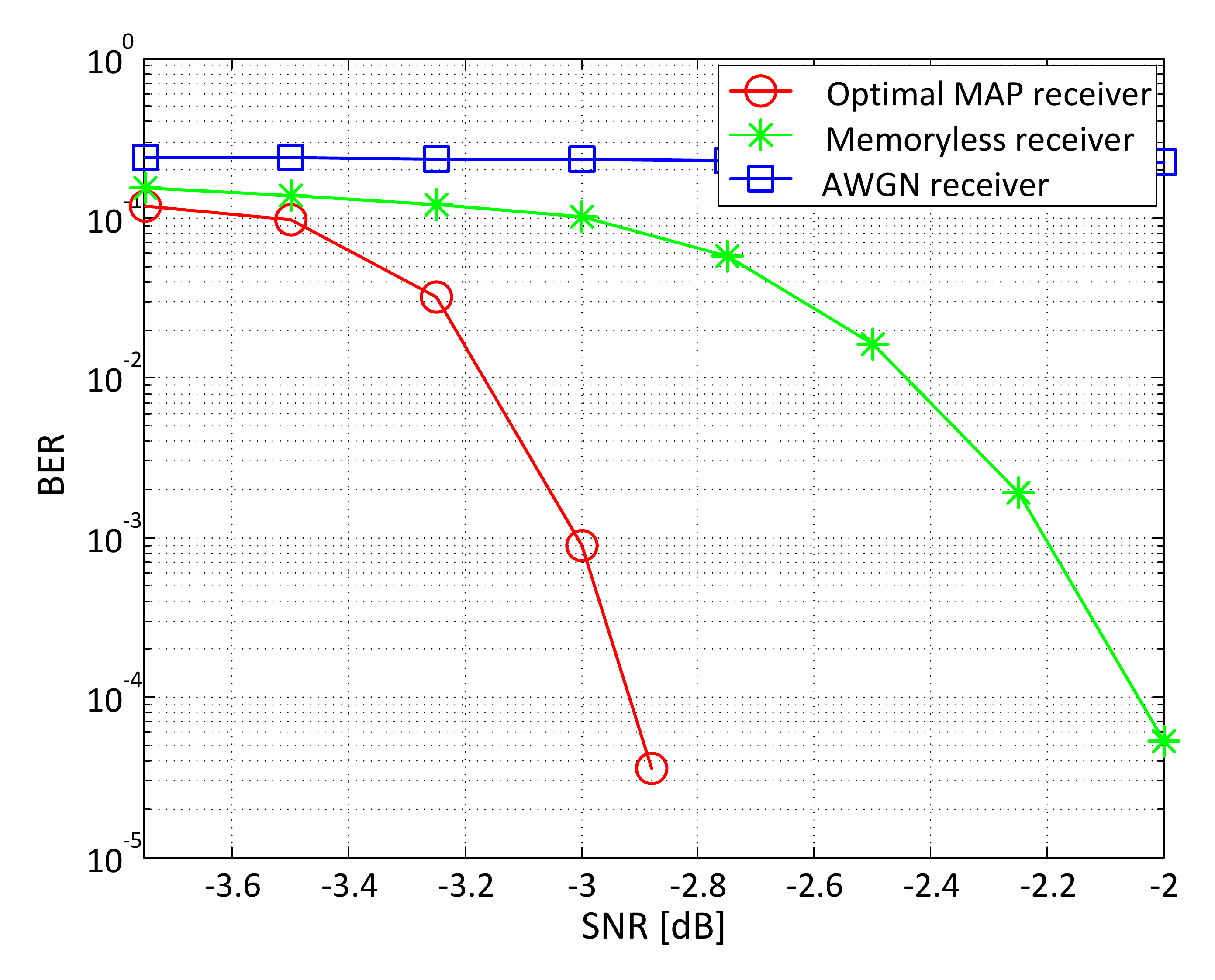}\\
  \caption{BER performances of coded selection decode-and-forward relaying (SDFR) scheme. A BPSK modulation is adopted and each channel is characterized by $p_B=0.1$, $\gamma=100$, $R=100$.}\label{ber_coded_scheme}
\end{figure}

Fig.~\ref{ber_selection_relaying} evaluates the analytical BER performances of selective DF cooperative communication system using (\ref{ber_SDF}) for different levels of threshold at the relay. As a performance benchmark, the performance of SR with optimal MAP receiver is also shown. From the numerical results, we observe that although in general SNR threshold-based selection relaying improves the BER performance compared to the simple relaying, but by utilizing the BER of the relay at the destination, the proposed MAP receiver-based simple relaying performs better than the SNR-based selection relaying regardless of the value of threshold at the relay. This again confirms the benefit of exploiting $\theta_m$ at the destination.
\begin{figure}[!t]
  \centering
  \includegraphics[scale=0.31]{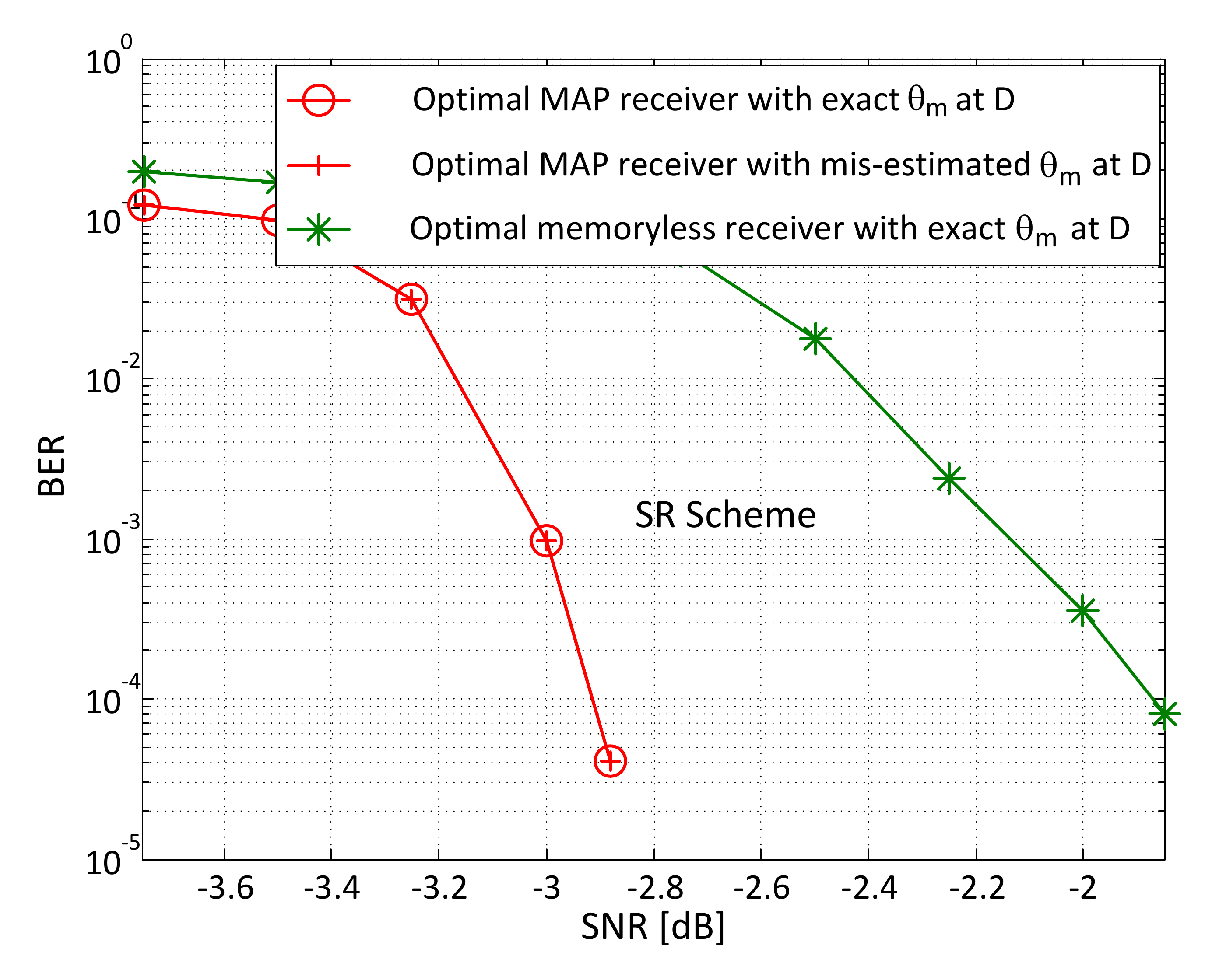}\\
  \caption{BER performances of coded simple relaying (SR) scheme assuming different realizations of $\theta_m$ at the destination. A BPSK modulation is adopted and each channel is characterized by $p_B=0.1$, $\gamma=100$, $R=100$.}\label{ber_coded_scheme_misestimation}
\end{figure}

We also considered systems employing powerful channel codes such as low-density parity check (LDPC) codes. Fig.~\ref{ber_coded_scheme} shows the simulated BER performances of SDFR scheme for LDPC coded transmission assuming three different detectors at the receiver side. It is assumed that a sequence of equally likely information bits of length $32,400$ is first encoded using LDPC channel coding based on the DVB-S2 standard with the code rate of $1/2$. The coded sequence is then interleaved using a random interleaver and mapped onto BPSK modulation sequence, and then transmitted over two state Markov-Gaussian channels each of which is characterized by $p_B=0.1$, $\gamma=100$, and $R=100$. For LDPC decoding at the relay as well as the destination, the number of iterations is set to $50$. As described earlier, the optimal MAP detector uses the MAP detection criterion, the memoryless detector is optimal for i.i.d. Bernoulli-Gaussian noise, and the AWGN detector is optimal for AWGN channel. As expected, from Fig.~\ref{ber_coded_scheme}, it is observed that similar to uncoded transmission, significant performance gains are achieved when the noise memory is taken into account in the detection process. Indeed, in the BER range of $10^{-5}$, the BER obtained with the memoryless receiver can be divided by almost $10^{3}$ to get the BER with optimal MAP receiver.
\begin{figure}[!t]
  \centering
  \includegraphics[scale=0.31]{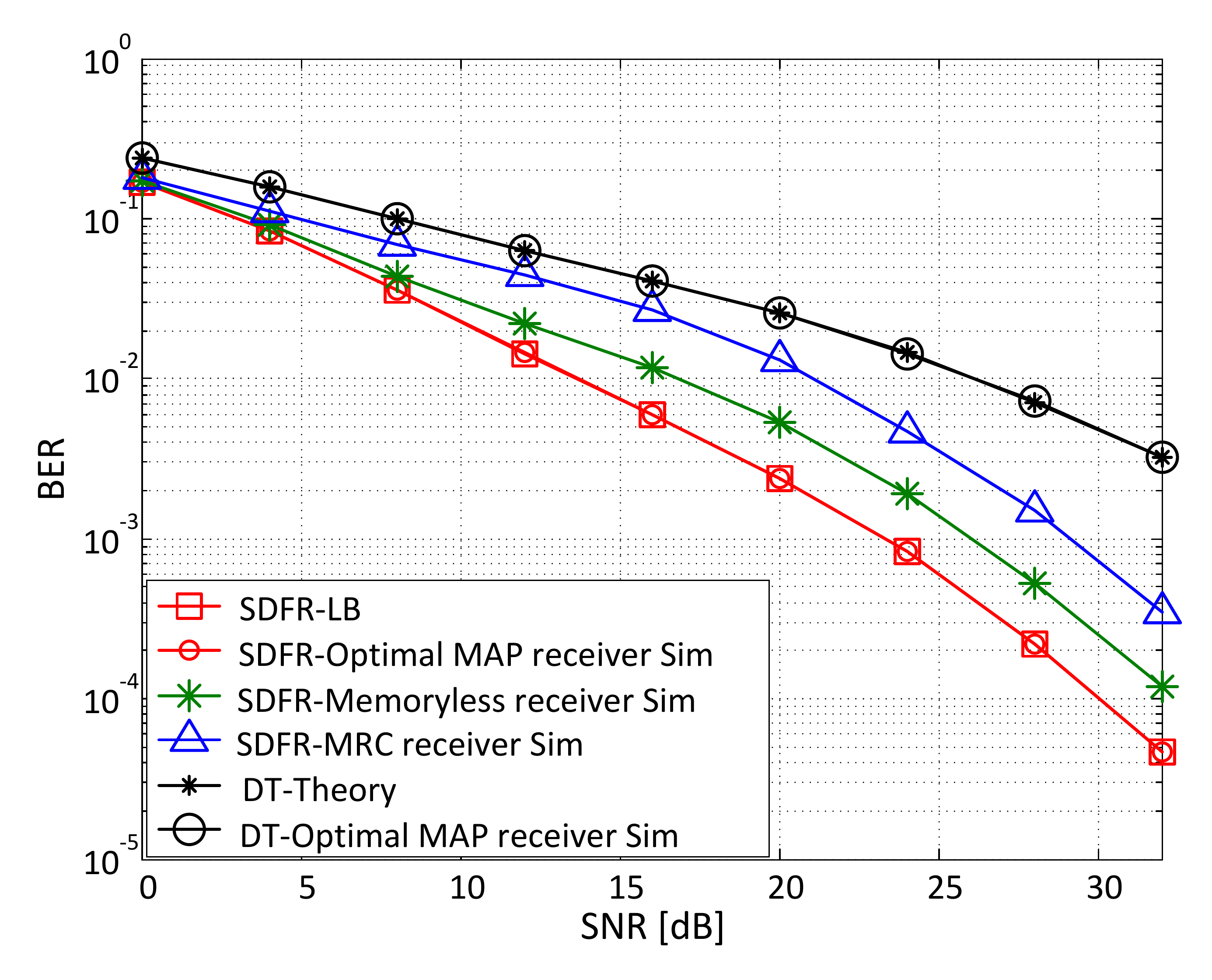}\\
  \caption{Analytical and simulated BER performances of direct transmission (DT) and selection decode-and-forward relaying (SDFR) schemes against SNR. A system employing a Q-PSK modulation is considered and the performance of various decoding schemes over two-state Markov-Gaussian channels, each characterized by $p_B=0.1$, $\gamma=100$, $R=100$ is shown.}\label{ber_SDFR_QPSK}
\end{figure}
\begin{figure}[!t]
  \centering
  \includegraphics[scale=0.31]{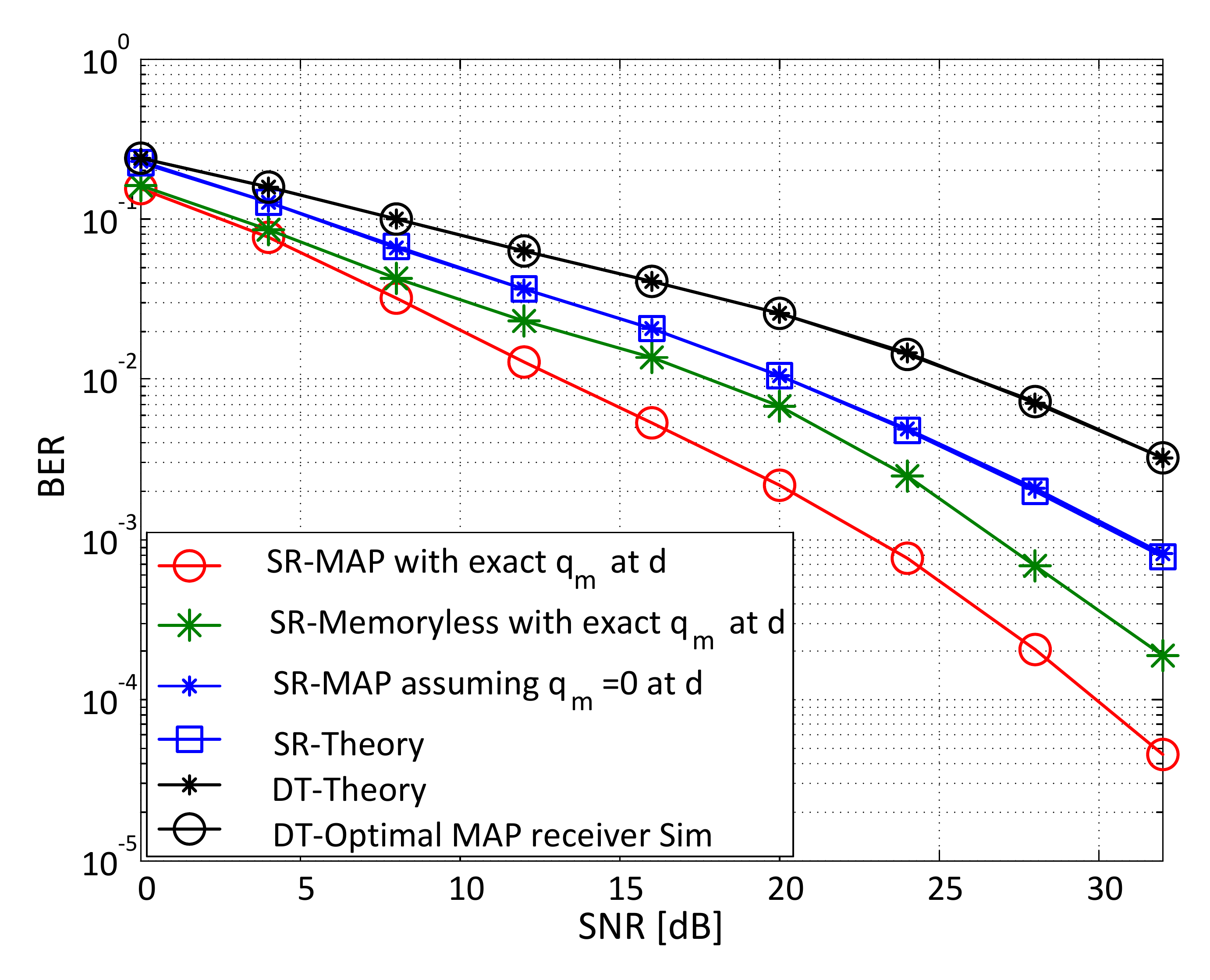}\\
  \caption{Analytical and simulated BER performances of direct transmission (DT) and simple relaying (SR) scheme against SNR with different realizations of $q_m$ at the destination. A Q-PSK modulation is adopted and each channel is characterized by $p_B=0.1$, $\gamma=100$, $R=100$.}\label{ber_SR_QPSK}
\end{figure}

Fig.~\ref{ber_coded_scheme_misestimation} also shows the performance of SR scheme with MAP receiver using the following two different realizations of $\theta_m$ at the destination in case of coded transmission: (i)- the destination has perfect knowledge about $\theta_m$ and is utilized in the detection process and, (ii)- $\theta_m$ is estimated at the destination with $10$ percent estimation error for utilization. It is obvious from Fig.~\ref{ber_coded_scheme_misestimation} that similar to SDFR, significant performance gains are achieved in SR scheme, when the noise memory is taken into account in the detection process. Interestingly, the performance gain is practically the same, even if the destination does not have perfect knowledge about $\theta_m$.

\begin{figure}[!t]
  \centering
  \includegraphics[scale=0.31]{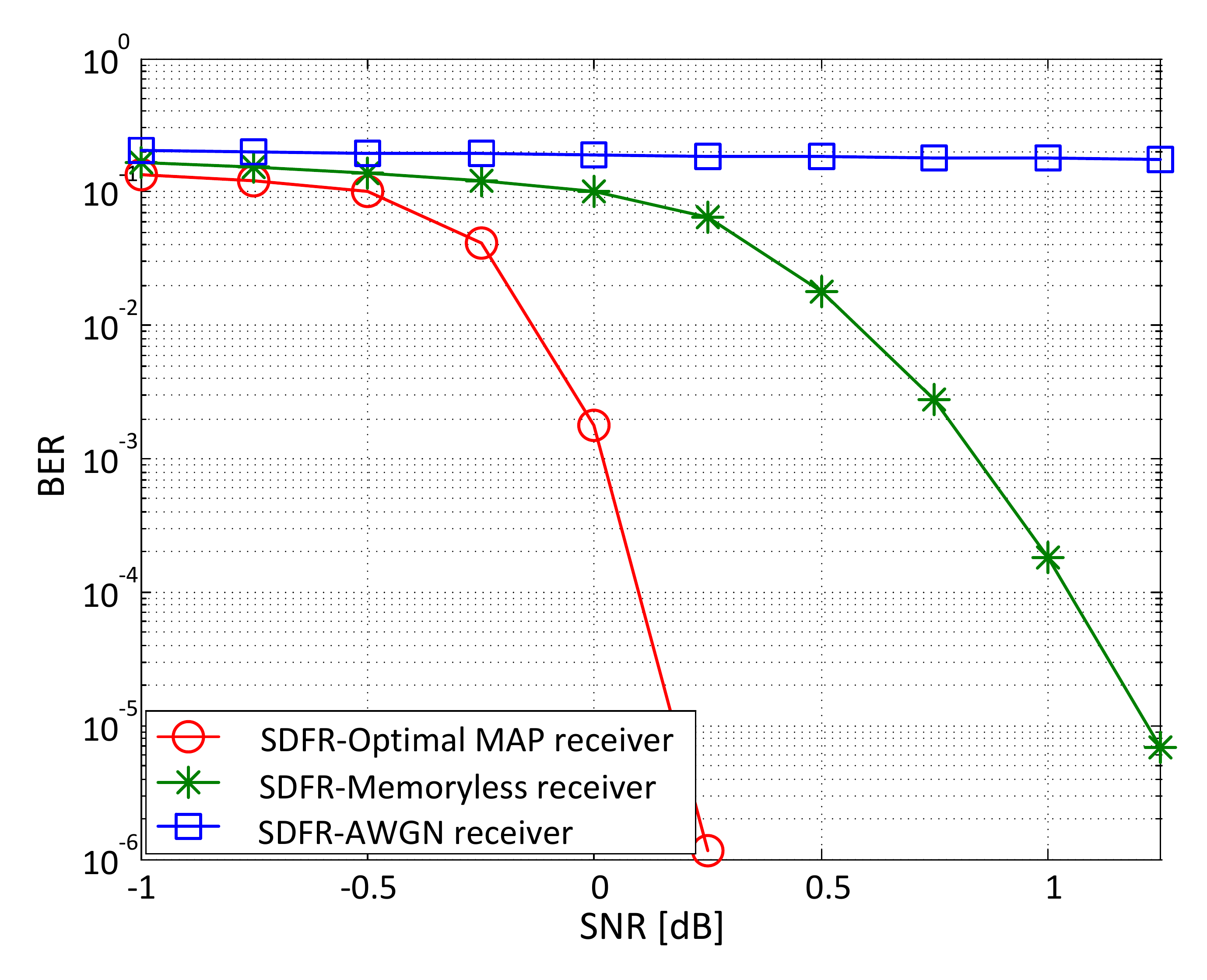}\\
  \caption{BER performances of coded selection decode-and-forward relaying (SDFR) scheme. A Q-PSK modulation is adopted and each channel is characterized by $p_B=0.1$, $\gamma=100$, $R=100$.}\label{ber_coded_SDFR_QPSK}
\end{figure}

So far, we have assumed BPSK modulation. Finally, we study the performances of DT and DF cooperative relaying schemes using Gray-coded Q-PSK modulation under both uncoded and coded scenario. We assume the same SNR for both BPSK and Q-PSK modulation schemes. The obtained results are shown in the figures from Fig.~\ref{ber_SDFR_QPSK} - Fig.~\ref{ber_coded_SDFR_QPSK}. From the obtained results it is seen that as in BPSK, the same arguments are hold for Q-PSK modulation scheme also, i.e., the analytical result matches well the simulation result. Also, the proposed MAP receiver attains the lower bound derived for DF CR scheme, and leads to large performance gains compared to the conventional receiving criteria which were optimized for additive white Gaussian noise (AWGN) channel and memoryless impulsive noise channel.

\section{Conclusion}\label{conclusion}
Cooperative relaying has been identified as a promising technology since last decade due to its reliability over fading and interference channels. In this article, we have presented the mathematical model to verify the analytical and simulated performances for DF CR schemes over time-correlated impulsive noise channel in the presence of Rayleigh fading. We also investigated the receiver structure at the destination for the proposed model. From the obtained results, it is observed that the analytical results agree with the simulations and our proposed MAP receiver achieves the lower bound derived for DF CR scheme, and performs significantly better than the conventional schemes developed for additive white Gaussian noise channel and memoryless impulsive noise channel. Also, DF CR scheme performs significantly better than DT under the same power consumption. Additionally, for simple relaying, the proposed MAP receiver achieves an SNR gain of around $8$ dB by utilizing the relay-induced BER at the destination and attains similar performance as obtained through selective DF relaying.

%
\section*{Acknowledgment}
This work was supported by Hydro-Quebec, the Natural Sciences and Engineering Research Council of Canada, and McGill University in the framework of the NSERC/Hydro-Quebec/McGill Industrial Research Chair in Interactive Information Infrastructure for the Power Grid.

\ifCLASSOPTIONcaptionsoff
  \newpage
\fi

\bibliographystyle{ieeetr}
\bibliography{book_ieee}

\end{document}